\documentclass[%
reprint,
superscriptaddress,
amsmath,amssymb,
aps,
prl,
]{revtex4-2}

\usepackage{graphicx}
\usepackage{dcolumn}
\usepackage{bm}
\usepackage{mathtools}
\usepackage{stackrel}
\usepackage[colorlinks=true,linkcolor=blue,citecolor=blue,filecolor=blue,urlcolor=blue]{hyperref}

\usepackage{siunitx}
\newcommand{\um}{\SI{}{\micro\metre}}

\begin{document}


\title{Impact of surface charge depletion on the free electron nonlinear response of heavily doped semiconductors}

\author{Federico De Luca}
 \email{federico.deluca@iit.it}
\affiliation{Istituto Italiano di Tecnologia, Center for Biomolecular Nanotechnologies, Via Barsanti 14, 73010 Arnesano, Italy}
\affiliation{Dipartimento di Matematica e Fisica "E. De Giorgi", Universit\`a del Salento, via Arnesano, 73100 Lecce, Italy}

\author{Cristian Cirac\`i}%
 \email{cristian.ciraci@iit.it}
\affiliation{Istituto Italiano di Tecnologia, Center for Biomolecular Nanotechnologies, Via Barsanti 14, 73010 Arnesano, Italy}

\date{\today}

\begin{abstract}
We propose surface modulation of the equilibrium charge density as a  technique to control and enhance, via an external static potential, the free electron nonlinear response of heavily doped semiconductors.
Within a hydrodynamic perturbative approach, we predict a two order of magnitude boost of free electron third-harmonic generation.
\end{abstract}

\maketitle

Among the main challenges of modern applied physics, the control and the concentration of light at the subwavelength scale are of extreme importance for the realization of integrated optical technologies, especially to reach operational efficiencies in devices based on nonlinear optical effects, which otherwise would require high laser intensities and long propagation distances in macroscopic nonlinear crystals. Toward the accomplishment of this purpose, the study of the coupling of light with the collective oscillation of free electrons (FEs) in materials characterized by a high density of such carriers, i.e. plasmonics, has a central role.  
Plasmonic nanoantennas have been commonly used as local-field amplifier
in hybrid systems to enhance optical nonlinearity from dielectric material placed in their vicinity \cite{Deng:2020gq, QixinShen:2020cw,Noor:2020kq,Sarma:2019im}, however, the nonlinear response may also arise directly from the plasmonic material itself, specifically from the dynamics of nonequilibrium FEs \cite{Scalora:2010kd, Ciraci:2012vk, Kauranen:2012ff, krasavin2018free, DeLuca:19, Rodriguez:2020s, noor:2022}.
Notoriously, noble metals are the main constituents of plasmonic devices in the visible spectrum. On the other hand, heavily doped semiconductors (i.e. with charge densities $n_0\sim 10^{19}$--$10^{20}$~cm$^{-3}$) have emerged as alternative materials for plasmonics in the near-infrared (NIR), i.e. $0.8<\lambda<2$~\um, and in the mid-infrared (MIR), i.e. $2<\lambda<20$~\um~\cite{Frigerio:Tunability:2016, Maack:2017fn, Fischer:2018bs, Fischer:21, Chavarin:21}. Being low-loss high-quality materials that can be compatible with standard microelectronics fabrication processes, and being their optical response tunable through electrical or optical doping, heavily doped semiconductors offer a unique perspective for integrated optical devices in the NIR and in the MIR \cite{soref2010mid, Boltasseva:2011ep, Naik:2013am,Taliercio:2019}. 

Within this context, we have recently investigated the FE nonlinear optical dynamics of heavily doped semiconductors, predicting that cascaded third-harmonic generation (THG) due to second-harmonic signals can be as strong as direct THG contributions, even when the second-harmonic generation efficiency is zero, and showing that, when coupled with plasmonic enhancement, FE nonlinearities could be up to two orders of magnitude larger than conventional semiconductor nonlinearities \cite{deluca2021, deluca:epj2022}. We employed a hydrodynamic description that includes terms up to the third order, usually negligible for noble metals. This choice has been made taking into consideration that, within the hydrodynamic formalism, the third-order response, expressed through the third-order polarization vector $\mathbf{P}^{(3)}_{\rm NL}$, is inversely proportional to the squared equilibrium charge density, i.e. $\mathbf{P}^{(3)}_{\rm NL} \propto \frac{1}{n_0^2}$. Indeed, doped semiconductors with a plasma wavelength in the MIR have a charge density ($n_0\sim10^{19}$~cm$^{-3}$) much lower than noble metals, such as gold ($n_0\sim10^{22}$~cm$^{-3}$). Hence, FE nonlinearites may grow as much as six orders of magnitude, overcoming by far the contributions originating in the crystal lattice nonlinear susceptibility $\chi^{(3)}$, which instead represents the dominant third-order nonlinear source in gold due to the high concentration of charge carriers. Moreover, the nonlinear active volumes are expected to increase in semiconductors due to to their smaller effective masses \cite{deluca2021}.

A further step forward, along this direction  can be made if another very important characteristic of hydrodynamic nonlinearities is considered: they emerge predominantly at the surface \cite{Sipe:1980vz, Ciraci:2012vw}. As a consequence, an induced decrease of the electron density, in a small region of the semiconductor very close to its surface, may be exploited to increase the nonlinear response strength of the plasmonic system. In doped semiconductors, such a modification of the charge density can be obtained through the application of an external bias, i.e. by means of field-effect modulation \cite{Zandi:2018jv, ghini2022control}. Therefore, this technique may provide the unique ability to externally and dynamically modulate the nonlinear coefficients of heavily doped semiconductors by a simple setting of DC electric potential levels. 
In this Letter we present a model for describing the influence of surface charge depletion on FE nonlinearities and make quantitative predictions about the role of field-effect modulation for the control of the optical nonlinear response of heavily doped semiconductors. 
Finally, we demonstrate a two order of magnitude enhancement in THG from a doped InP grating.

As in our previous works on FE nonlinearities \cite{deluca2021, deluca:epj2022}, for the representation of nonlinear and nonlocal FE dynamics, we apply the quasi-classical formalism of the hydrodynamic model in the limit of Thomas-Fermi approximation \cite{Ciraci:2013dz, Ciraci:2017bp, Ciraci:2016il}.  Within this framework, the following constitutive relation is employed to model the FE fluid via two macroscopic variables, its charge density $n(\mathbf{r},t)$, and its current density $\mathbf{J}(\mathbf{r},t) = -en\mathbf{v}$, with $\mathbf{v}(\mathbf{r},t)$ being the electron velocity field: 
\begin{equation}
\label{eqn:HM}
\begin{split}
    \dot{\mathbf{J}} + \gamma\mathbf{J} &=
    \frac{e^2 n}{m}\mathbf{E} 
    -\frac{\mu_0 e}{n}\mathbf{J}\times\mathbf{H}
\\ 
    &+\frac{1}{e}\left(\frac{\mathbf{J}}{n} \nabla\cdot\mathbf{J}
    -\mathbf{J}\cdot\nabla\frac{\mathbf{J}}{n}\right)
    +\frac{e n}{m}\nabla \frac{\delta T[n]}{\delta n},
\end{split}
\end{equation}
where time derivatives are expressed in dot notation, $m$ is the electron effective mass, $e$ the elementary charge (in absolute value), $\mu_0$  is the magnetic permeability of vacuum and $\gamma$ is the damping rate. 
This equation portrays the many-body nonlinear dynamics of the charge carriers under the influence of external electric $\mathbf{E}(\mathbf{r},t)$ and magnetic $\mathbf{H}(\mathbf{r},t)$ fields. Furthermore, the fermionic nature of FEs, which cannot be compressed in an infinitesimally thin layer, is accounted by means of the electron pressure term, where $T[n]$,
is the kinetic energy functional. On the other hand, we neglect electron spill-out and apply hard-wall boundary conditions.

Employing a perturbative approach, we can write the charge density as a sum of a static and a dynamic term:
\begin{equation}
    n({\bf r},t)=n_0 (\mathbf{r})+n_{\rm d}({\bf r},t),
    \label{eqn:n}
\end{equation}
where $n_0(\mathbf{r})$ is the equilibrium state nonperturbed electron charge density and $n_{\rm d} \ll n_0 $ is the induced charge density, representing perturbative corrections to the equilibrium density. 
Similarly, the electric field and the kinetic functional can be written as 
$\mathbf{E}(\mathbf{r},t) = \mathbf{E}_0(\mathbf{r}) + \mathbf{E}_d(\mathbf{r},t)$ and 
$ T[n](\mathbf{r},t) = T_0[n_0(\mathbf{r})] + T_d[n(\mathbf{r},t)]$, respectively.
As a consequence, Eq.~\eqref{eqn:HM} can be split into a static and a dynamic equation:
\begin{subequations}
    \label{eqn:HM_split}
    \begin{eqnarray}
        \label{eqn:static}
        &&\nabla\frac{\delta T_0[n_0]}{\delta n_0} + e\mathbf{E}_0 = 0
    \\
        \label{eqn:dynamic}
        &&\dot{\mathbf{J}} + \gamma\mathbf{J}=
        \frac{e^2n}{m}\mathbf{E}_d
        -\frac{\mu_0 e}{n}\mathbf{J}\times\mathbf{H}
        \nonumber
    \\ 
        &&\hspace{0.9cm}
        +\frac{1}{e}\left(\frac{\mathbf{J}}{n} \nabla\cdot\mathbf{J}
        -\mathbf{J}\cdot\nabla\frac{\mathbf{J}}{n}\right)
        +\frac{e n}{m}\nabla \frac{\delta T_d[n]}{\delta n}.
    \end{eqnarray}
\end{subequations}
Eq.~\eqref{eqn:static} coupled to the Poisson equation would give a self-consistent expression for the equilibrium density and the static electric field. 
However, we calculate $n_0(\mathbf{r})$ by means of the method described in the Supplemental Material \cite{SM}, which takes into account bands bending in doped semiconductors within the parabolic band approximation \cite{Seiwatz:1958iq, Zandi:2018jv, ghini2022control}.
Note that this method is equivalent to solving Eq.~\eqref{eqn:static} for a proper expression of the static kinetic functional $T_0[n_0]$.
To solve the dynamic Eq.~\eqref{eqn:dynamic}, we consider the kinetic energy functional within the Thomas-Fermi approximation, 
i.e. $\frac{\delta T_d[n]}{\delta n}=\frac{5}{3}c_{\rm TF}\left(n^{\frac{2}{3}}-n_0^{\frac{2}{3}}\right)$,
with $c_{\rm TF} = \frac{\hbar^2}{m}\frac{3}{10}(3\pi^2)^{2/3}$.
Considering a Taylor expansion up to the third order, we can rewrite $n^{\frac{2}{3}}-n_0^{\frac{2}{3}}=n_0^{2/3} \left[\frac{2}{3}\frac{n_{d}}{n_0} 
    -\frac{1}{9}\left(\frac{n_{d}}{n_0}\right)^2 +\frac{4}{81}\left(\frac{n_{d}}{n_0}\right)^3 \right]$ 
such that, after some algebra, the quantum pressure term becomes:
\begin{eqnarray}
    &&\frac{e n}{m}\nabla \frac{\delta T_d[n]}{\delta n} \simeq
    e\beta^2\left[1+\frac{2}{3}\frac{n_{d}}{n_0}
    -\frac{1}{9}\left(\frac{n_{d}}{n_0}\right)^2\right]\nabla n_d
    \nonumber
    \\
    &&+ e\beta^2\left[-\frac{1}{3}\frac{n_{d}}{n_0}
    -\frac{1}{9}\left(\frac{n_{d}}{n_0}\right)^2 
    +\frac{4}{81}\left(\frac{n_{d}}{n_0}\right)^3 \right]\nabla n_0   
\end{eqnarray}
where $\beta(\mathbf{r})^2=\frac{10}{9}\frac{c_{\rm TF}}{m}n_0(\mathbf{r})^{2/3}$.
Eq.~\eqref{eqn:dynamic} can be then written in terms of the polarization field $\mathbf{P}(\mathbf{r},t)$, with $\dot{\mathbf{P}}=\mathbf{J}$, $n_{\rm d}=\frac{1}{e}\nabla \cdot{\bf P}$, and $n^{-1} \simeq n_0^{-1}\left(1-\frac{n_{d}}{n_0}\right)$, as:
\begin{equation}
\label{eqn:HM_P}
    \ddot{\mathbf{P}}+\gamma\dot{\mathbf{P}} = 
    \frac{n_0e^2}{m}{\mathbf{E}} 
    +\beta^2\nabla(\nabla\cdot\mathbf{P})
    -\frac{1}{3}\frac{\beta^2}{n_0}(\nabla\cdot\mathbf{P})\nabla n_0
    +\mathbf{S}_{\rm NL}. 
\end{equation}
where $\mathbf{S}_{\rm NL} = \mathbf{S}_{\rm NL}^{(2)} + \mathbf{S}_{\rm NL}^{(3)}$ includes second- and the third-order nonlinear sources, respectively:
\begin{subequations}
\label{eqn:NL_sources}
    \begin{eqnarray}
    \label{eqn:second_order}
        \mathbf{S}^{(2)}_{NL}=&&
        \frac{e}{m}\mathbf{E}\nabla\cdot{\mathbf{P}}
        -\frac{e\mu_0}{m}\dot{\mathbf{P}} \times {\mathbf{H}}
        \nonumber
    \\
        &&+\frac{1}{{e{n_0}}}({\dot{\mathbf{P}}\nabla  \cdot \dot{\mathbf{P}}
         +\dot{\mathbf{P}} \cdot \nabla \dot{\mathbf{P}}})
        -\frac{1}{en_0^2}\dot{\mathbf{P}}(\dot{\mathbf{P}}\cdot\nabla n_0)
        \nonumber    
    \\
        &&+\frac{1}{3}\frac{\beta^2}{e n_0}\nabla (\nabla\cdot\mathbf{P})^2
        -\frac{1}{9}\frac{\beta^2}{e n_0^2}(\nabla\cdot\mathbf{P})^2 \nabla n_0,
    \label{eqn:2 order}
    \end{eqnarray}
    \begin{eqnarray}
    \label{third_order}
        \mathbf{S}^{(3)}_{NL}=&&
        -\frac{1}{{{e^2}n_0^2}}
        \Big[
        \nabla\cdot \mathbf{P}(\dot{\mathbf{P}}\nabla  \cdot \dot{\mathbf{P}}
        +\dot{\mathbf{P}} \cdot \nabla \dot{\mathbf{P}}) 
        +\dot{\mathbf{P}} \cdot \dot{\mathbf{P}}\nabla\nabla \cdot {\mathbf{P}}
        \Big]\nonumber
    \\
        &&+\frac{2}{e^2n_0^3}(\nabla\cdot\mathbf{P})\dot{\mathbf{P}}(\dot{\mathbf{P}}\cdot\nabla n_0)
        \nonumber
    \\
        &&-\frac{1}{27}\frac{\beta^2}{e^2 n_0^2}\nabla(\nabla\cdot\mathbf{P})^3
        +\frac{4}{81}\frac{\beta^2}{e^2 n_0^3}(\nabla\cdot\mathbf{P})^3\nabla n_0.
    \label{eqn:3 order}
    \end{eqnarray}
\end{subequations}
A development with respect previous works \cite{deluca2021, Scalora:2010kd, Ciraci:2012vk} is represented by the introduction of nonlinear contributions proportional to $\nabla n_0$, by means of which we tackle the non-zero gradient of the equilibrium density. 
The aforementioned terms and all the surface contributions, i.e. those proportional to $\nabla \cdot \mathbf{P}$, describe nonlinear effects whose origin is at the surface of the material. Consequently, hydrodynamic nonlinearities are expected to be extremely sensitive to the changes of the physical condition at the surface, such as a change in the density $n_0$.

At this point, if a time-harmonic dependence of the fields is assumed, 
i.e. ${\bf{F}}({\bf{r}},t) = \sum\limits_j {{{\bf{F}}_j}({\bf{r}}){e^{-i{\omega_j}t}}}$, 
with ${\bf{F}}={\bf{E}}$, ${\bf{H}}$, or ${\bf{P}}$,
combining Eqs.~\eqref{eqn:HM_P} and \eqref{eqn:NL_sources} with Maxwell's equations, the following system can be derived for each harmonic $\omega_j$:
\begin{subequations}
\label{eqn:HM_set}
    \begin{eqnarray}
        &&\nabla\times\nabla\times \mathbf{E}_j
        -\varepsilon\frac{\omega_j^2}{c^2}\mathbf{E}_j
        -\omega_1^2\mu_0 (\mathbf{P}_j+\mathbf{P}_{\omega_j}^{\rm NL})=0,    
        \label{eqn:maxwell}
\\
        &&\beta^2\nabla ({\nabla\cdot\mathbf{P}_j})
        -\frac{1}{3}\frac{\beta^2}{ n_0}(\nabla\cdot\mathbf{P}_j)\nabla n_0
        +(\omega^2+i\gamma\omega)\mathbf{P}_j
        \nonumber
    \\
        &&= -\frac{n_0e^2}{m}\mathbf{E}_j +\mathbf{S}_{\omega_j},
        \label{eqn:HM_harmonic}
    \end{eqnarray}
\end{subequations}
where local contributions from the semiconductor, both linear, through the local permittivity $\varepsilon$, and nonlinear, through the nonlinear polarization $\mathbf{P}_{\omega_j}^{\rm NL}$ are considered.
Since a coupling between different harmonics occurs through the nonlinear contributions $\mathbf{P}_{\omega_j}^{\rm NL}$ and $\mathbf{S}_{\omega_j}$, Eqs.~\eqref{eqn:HM_set} constitute a set of coupled nonlinear differential equations, whose resolution is not straightforward. For this reason, as we expect harmonic signals to be several orders of magnitude smaller than the pump fields, we assume that the latter is not affected by the nonlinear process (undepleted pump approximation), i.e. $\mathbf{P}_{\omega_1}^{\rm NL}=\mathbf{S}_{\omega_1}=0$.
The system of Eqs.~\eqref{eqn:HM_set} reduces then to separated sets of one-way coupled equations, one for each harmonic.
Moreover, since our goal is to study the impact of surface depletion on FE nonlinearities, we neglect contribution from the background lattice, i.e. $\mathbf{P}_{\omega_3}^{\rm NL} = 0$.
In what follows, we focus on FE THG, both direct, i.e. a third-order process where three photons of energy $\hbar\omega$ combine to give a single photon of energy $3\hbar\omega$, and cascaded, i.e. a combination of two second-order processes, namely second-harmonic generation (SHG) and sum-frequency generation. The corresponding expressions of the nonlinear sources, $\mathbf{S}_{\omega_j}$, derived from Eqs.\eqref{eqn:NL_sources}, are reported in the Supplemental Material \cite{SM}.

\begin{figure}[t]
	\centering
	\includegraphics[width=8.6 cm]{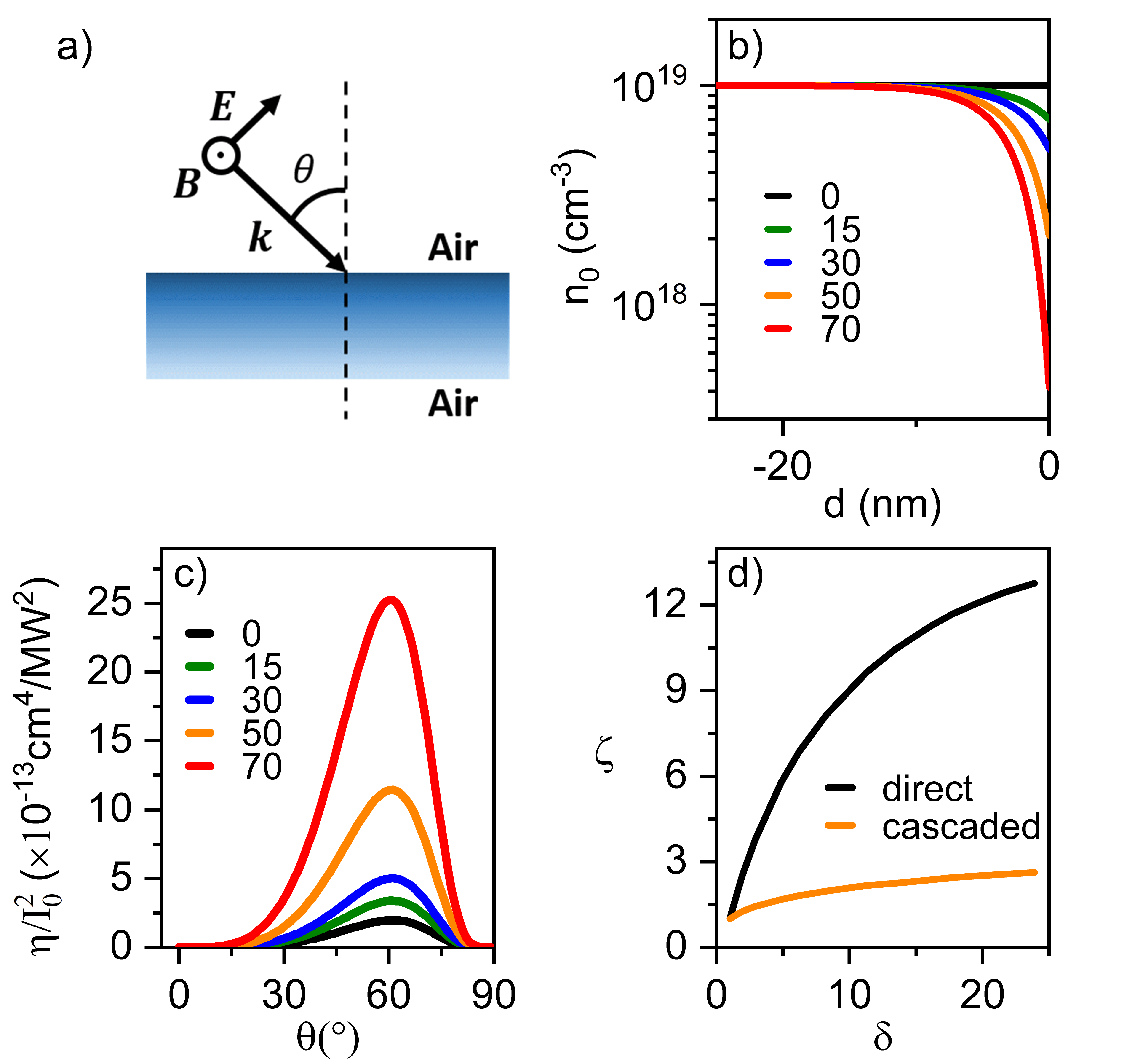}
	    \caption{Effects of surface charge depletion on the FE THG efficiency $\eta$ of a doped InP slab:
	    (a) a TM plane wave impinging on a semi-infinite geometry is considered;
	    (b) equilibrium charge density $n_0 (\mathbf{r})$ as a function of the distance $d$ from the surface of the slab for different levels of modulation (in V/\um); 
	    (c) related $\eta$, in the case of direct THG, normalized to the squared input intensity $I_0^2$, as a function of the angle of incidence $\theta$;
	    (d) comparison of the enhancement factors, $\zeta$, of direct and cascaded FE THG, as a function of the depletion factor, $\delta$, in correspondence of the peak efficiencies at $\theta = 60$°. 
	    }
	\label{fig:slab}
\end{figure} 
In order to estimate the impact of surface charge depletion on the FE nonlinear response of heavily doped semiconductors, we first apply the developed formalism to calculate the FE THG efficiency $\eta$ of a semiconductor slab. 
In particular, we solved Eqs.~\eqref{eqn:HM_set} numerically using the finite-elements method within a customized frequency-dependent two dimensional implementation in COMSOL Multiphysics \cite{comsol}.
The efficiency has been calculated by normalizing the power of the generated signal to the input power at the fundamental frequency, $\eta=I_{\rm G}/I_0$, where $I_{\rm G}$ is the generated intensity. As a consequence, for third-order nonlinearities, $\eta$ will scale with $I_0^2$.

To accurately model the semiconductor's linear response, on top of the Drude-like dispersion described by Eq.\eqref{eqn:HM_P},  we consider a local permittivity contribution, $\varepsilon_\infty$, such that, neglecting nonlocal effects, we retrieve the usual dielectric function $\varepsilon (\omega)= \varepsilon_\infty-\frac{\omega_P^2}{\omega^2+i\gamma\omega}$, where $\omega_P=\sqrt{\frac{e^2 n_0}{\varepsilon_0 m}}$ is the plasma frequency of the semiconductor, being $\varepsilon_0$ the dielectric constant of vacuum. The input field is a TM plane wave impinging on the geometry with a certain angle of incidence $\theta$ (see Fig.~\ref{fig:slab}(a)).
The slab is characterized by an equilibrium charge density profile modulated in a very small region at the top interface, as shown by the curves in Fig.~\ref{fig:slab}(b), calculated as described in the Supplemental Material \cite{SM}. The imposed boundary conditions on the top surface of the semiconductor slab correspond to an applied static electric field that can be up to $70$~V/\um. 

The material considered for this work is Indium Phosphide (InP), a direct bandgap III-V semiconductor and a low loss plasmonic material for the MIR region \cite{Naik:2013am, Panah:2016InP, Panah:2017s}. InP is, thanks to its intrinsic properties ($m=0.078~m_e$, $\varepsilon_{\infty}=9.55$ \cite{Naik:2013am}), among the most promising semiconductors for the realization of integrated optical platforms based on FE nonlinear dynamics \cite{deluca:epj2022}.
Since we assume the value of the equilibrium charge density in the bulk to be $n_b = 10^{19}$~cm$^{-3}$, the simulated InP's slab  has a screened plasma wavelength in the MIR, $\tilde{\lambda}_{\rm p}=9.1$~\um, where $\tilde{\lambda}_{\rm p}=\frac{2\pi c}{\tilde{\omega}_{\rm p}}$, with $\tilde{\omega}_{\rm p}= {\omega}_{\rm p}/\sqrt{\varepsilon_\infty}$ being the screened plasma frequency.
Finally, $\gamma = 1$~ps$^{-1}$ has been assumed dispersion-less \cite{deluca2021}. 
Note that, given the dimension of the system, the effects of the depletion region on the linear properties of the semiconductor are not sensitive.

In Fig.~\ref{fig:slab}(c), we report $\eta$ in the case of direct THG, normalized to the squared input intensity $I_0^2$, as a function of $\theta$ for the five different $n_0 (\mathbf{r})$ profiles of plot b, at a fundamental field (FF) wavelength $\lambda_{\rm FF} = 12$~\um, while, in Fig.~\ref{fig:slab}(d), we compare the enhancement factors $\zeta=\eta/\eta_0$ (where $\eta_0$ is the THG efficiency obtained with no applied potential) of direct and cascaded THG, in correspondence of the peak efficiencies (i.e. for $\theta = 60$°), as a function of the depletion factor $\delta = n_b/{n_0^{\rm surf}}$, where $n_0^{\rm surf}$ is the value of $n_0$ for $d=0$.
Here, the angular dispersion of $\eta$ is that peculiar of third-order FE THG, i.e. it is null at normal incidence and grows with $\theta$, peaking at a high angle of incidence. The reason is that, for $\theta=0$, the electric field is parallel to the slab, as a result there cannot be charge oscillations of the charge carriers in the finite dimension of the slab \cite{deluca2021}. 
Instead, the important feature emerging from Fig.~\ref{fig:slab} is the boost of FE THG because of the localized diminution of $n_0 (\mathbf{r})$ in a very thin region ($\sim 10$~nm) in proximity of the surface of the doped semiconductor. Indeed, as it can be observed more clearly in Fig.~\ref{fig:slab}(d), the enhancement factor of direct THG can be larger than one order of magnitude for $\delta \approx 25$, i.e. for $n_0^{\rm surf}$ about 25 times smaller than $n_b$. Conversely, with the same condition, the enhancement reached for the cascaded THG is $\zeta \approx 3$. To understand this result it should be observed that, notwithstanding the fact that the efficiency of SHG is approximately constant with $\delta$, the SH field in a region close to surface of the slab grows increasing the modulation. It follows that, the SFG, and consequently the cascaded THG, are enhanced because of the depletion.   

\begin{figure}[t]
	\centering
	\includegraphics[width=8.6 cm]{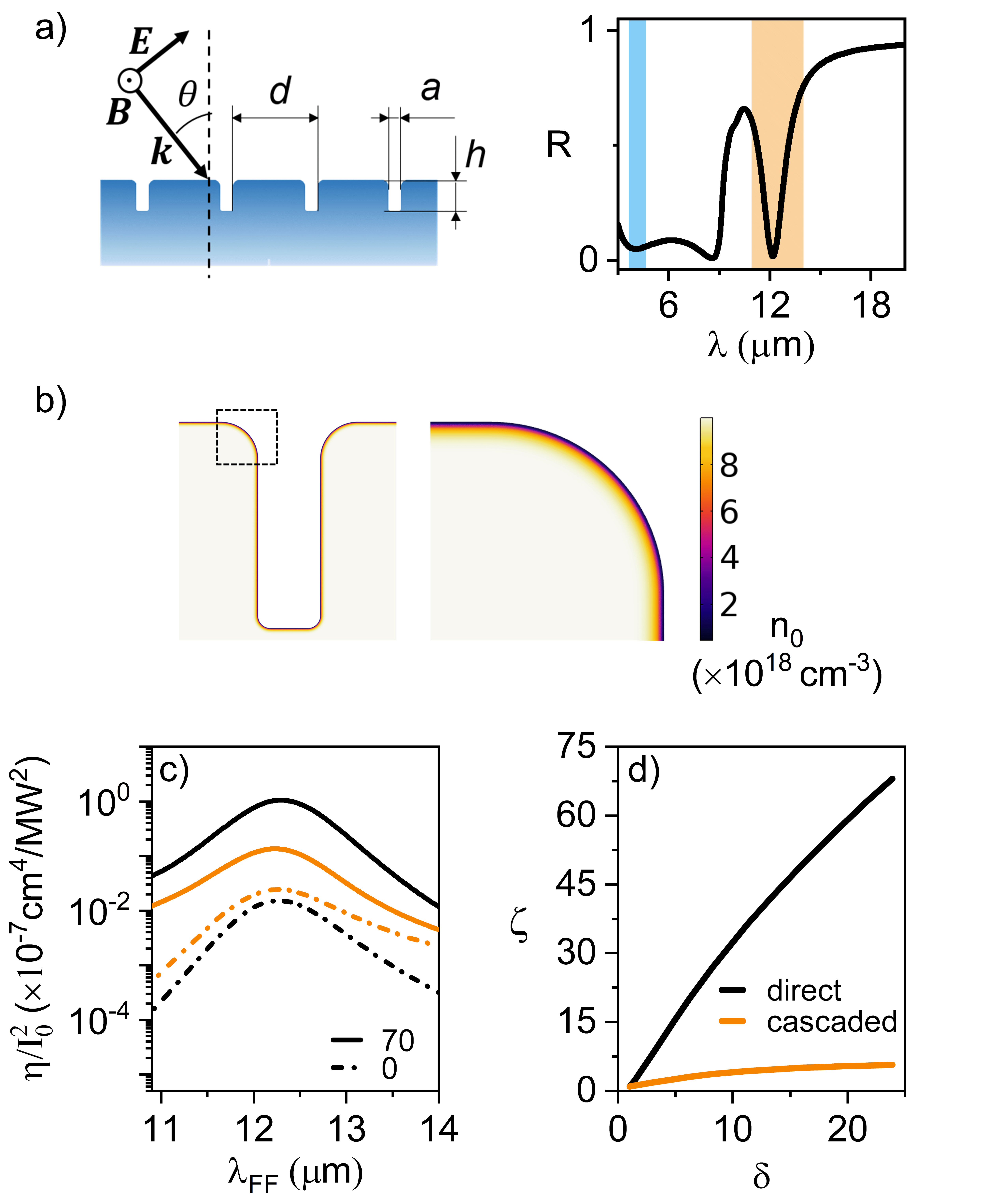}
	    \caption{Effects of surface charge depletion on the FE THG efficiency $\eta$ of a doped InP semi-infinite grating: 
	    (a) the structure and its reflectance R at normal incidence for $n_b = 10^{19}$~cm$^{-3}$, $d = 1$~\um, $a = 150$~nm, $h = 500$~nm. The orange and the light blue shadows evidence the considered range of variation of $\lambda_{\rm FF}$ and that of the corresponding $\lambda_{\rm TH}$, respectively. 
	    (b) $n_0 (\mathbf{r})$ along the grooves contour in the case of maximum depletion of Fig.~\ref{fig:slab}(b); 
	    (c) normalized efficiencies of direct and cascaded THG as a function of $\lambda_{\rm FF}$ in the proximity of the resonance, at normal incidence, in the case of zero and maximum modulation (in V/\um);  
	    (d) related enhancement factors, $\zeta$, as a function of the depletion factor, $\delta$, in correspondence of the peak efficiencies at $\lambda_{\rm FF} = 12.2$~\um.
	    }
	\label{fig:grooves}
\end{figure} 
As a next step, it may be interesting to employ our hydrodynamic formalism with the aim of studying the possible impact of charge depletion on the nonlinear response of a nanopatterned semiconductor slab characterized by a localized plasmon resonance in the MIR. 
We consider an infinite array of subwavelength grooves (a grating) portrayed in Fig.~\ref{fig:grooves}(a), a structure that supports plasmonic resonances and allows to couple virtually all incident energy into the active material at normal incidence and locally enhance the pump field \cite{Dechaux:2016hg}.
The pattern has been designed, as a function of the parameter $a$, $h$, and $d$, in order to be resonant in the MIR for a TM-polarized excitation, obtaining a resonance around $\lambda_{\rm FF} = 12.2~$\um, i.e. where the reflectance is almost zero. The doped semiconductor and the boundary conditions are the same considered for the simple slab. The difference is that now the region of charge depletion follows the contour of the grooves, as depicted in Fig.~\ref{fig:grooves}(b).
To study the nonlinear properties of the grating, we report in Fig.~\ref{fig:grooves}(c) the normalized efficiencies of direct and cascade THG, in this case as a function of $\lambda_{\rm FF}$, in proximity of the resonance and at normal incidence, showing a comparison of the undepleted cases with those of maximum modulation. 
In all cases, the maximum efficiency reached is about five order of magnitude larger that that obtained for the simple slab.
Finally, in Fig.~\ref{fig:grooves}(d), we portray $\zeta$ in correspondence of the peak efficiencies of plot (c), as a function of the depletion factor. 
Here, an enhancement of the efficiency when the depth of the region of depletion increases can be put in evidence also for the grating. Nevertheless, in Fig.~\ref{fig:grooves}(d), $\zeta$ is in all cases larger if compared to the same points in Fig.~\ref{fig:slab}(d), reaching $\zeta\approx 70$ (direct THG) and $\zeta\approx 6$ (cascaded THG) for $\delta\approx 25$. The peak efficiency of direct THG is larger than $10^{-5}$ if an input intensity of $10~\rm{MW/cm}^2$ is assumed. 

In conclusion, in order to evaluate the impact of surface charge depletion on the FE nonlinear response of heavily doped semiconductors, we have introduced a hydrodynamic perturbative approach that takes into account the non-zero gradient of $n_0 (\mathbf{r})$. We have employed our method to study THG in a simple slab and in a resonant grating of doped InP, showing a boost of the efficiency of generation caused by the localized diminution of $n_0 (\mathbf{r})$ on the surface of the material, and predicting an enhancement of the THG up to two order of magnitude with an applied external static bias of $70$~V/\um.
Our work highlights the role of field-effect gated modulation as a groundbreaking tool to externally and dynamically control the nonlinear coefficients of heavily doped semiconductors, opening a new route toward the development of integrated nonlinear optics at MIR frequencies.

\bibliography{library}

\begin{widetext}
\section{Supplemental Material}
\subsection{Space varying equilibrium charge density}
We derive the surface depleted equilibrium charge density, $n_0 (\mathbf{r})$, of the heavily doped semiconductor, following the approach developed by Seiwatz and Green \cite{Seiwatz:1958iq} and recently applied in Refs. \cite{Zandi:2018jv, ghini2022control}. Specifically, we solve the following dimensionless Poisson's equation:
\begin{equation}
\label{eqn:Poisson}
    \nabla^2 u = -\frac{e^2 n_0}{\varepsilon \varepsilon_0 k T},
\end{equation} 	
where $u(\mathbf{r})= \frac{E_F(\mathbf{r})-E_I}{kT}$ is a non-dimensional potential expressing the difference between the neutral bulk and the surface potentials, with $E_F$ being the flat band Fermi level and $E_I$ the center of the band gap, while $e$ is the elementary charge (in absolute value), $\varepsilon$ the static permittivity of the semiconductor, $\varepsilon_0$ the dielectric constant of vacuum, $k$ the Boltzmann constant and $T$ is the temperature. 

In general, $n_0 =\rho_D-\rho_A+\rho_p-\rho_n$, where $\rho_D$ is the donor dopant density, $\rho_A$ the acceptor dopant density, $\rho_p$ the hole density, and $\rho_n$ the electron density. Here, since we are studying a n-type semiconductor, we assume $\rho_A = 0$ (i.e. there are only donor dopants), while, within the parabolic band approximation, the other quantities are:
\begin{equation}
    \begin{split}
        &\rho_D = \frac{N_D}{1 + 2e^{(u-w_{D,I})}},
    \\
        &\rho_n= 4 \pi \left(\frac{2mkT}{h^2}\right)^{\frac{3}{2}} F_{\frac{1}{2}}(u-w_{CB,I}),
    \\
        &\rho_p= 4 \pi \left(\frac{2m_hkT}{h^2}\right)^{\frac{3}{2}} F_{\frac{1}{2}}(w_{VB,I}-u),
    \end{split}
\end{equation}
where 
$F_{\frac{1}{2}}(\eta)=\int_{0}^{\infty}\frac{x^{\frac{1}{2}}dx}{1+e^{(x-\eta)}}$ is a Fermi-Dirac integral, $N_D$ is the donor concentration, $m_h$ is the hole effective mass, and the quantities
$w_{i,I} = \frac{E_i-E_I}{kT}$, 
with $i= D, CB, VB$, depends on the donor (D) level $E_D$, on the conduction band (CB) minimum $E_{CB}$ and on the valence band (VB) maximum $E_{VB}$, respectively.
Eq.~\eqref{eqn:Poisson} then becomes:
\begin{equation}
\label{eqn:Poisson2}
    \begin{split}
        \nabla^2 u = -\frac{e^2}{\varepsilon \varepsilon_0 k T}\bigg[\frac{N_D}{1 + 2e^{(u-w_{D,I})}}
        +4 \pi \left(\frac{2mkT}{h^2}\right)^{\frac{3}{2}} F_{\frac{1}{2}}(u-w_{CB,I})
        -4 \pi \left(\frac{2m_hkT}{h^2}\right)^{\frac{3}{2}} F_{\frac{1}{2}}(w_{VB,I}-u)\bigg]. 
    \end{split}
\end{equation}

In this equation, the application of a static electric field $\mathbf{E}_0$ at the surface of the semiconductor is translated into the boundary condition $u = u_{\rm surf}=\frac{E_{\rm surf}-E_I}{kT}$. 
Once $u(\mathbf{r})$ is derived from Eq.~\eqref{eqn:Poisson2}, $n_0(\mathbf{r})$ can be calculated from Eq.~\eqref{eqn:Poisson} and implemented into the hydrodynamic equations, instead $\mathbf{E}_0 =-\frac{kT}{e}\nabla u$.

We solved Eq.~\eqref{eqn:Poisson2} numerically, with the finite-elements method, using the built in \textit{Poisson's Equation} module in COMSOL Multiphysics \cite{comsol}.
The parameters employed for the calculation are reported in Table~\ref{table:parameters}, while the properties of the semiconductor studied, indium phosphide (InP), are in Table~\ref{table:InP_parameters}. Our convention is to consider $E_{VB}$ as the zero-potential level and to write all the other levels, including the condition on the surface $E_{\rm surf}$, as a multiple of the energy gap $E_g$. With the assumed values of $E_D$ and $N_D$, the obtained flat band Fermi level is $E_F= 1.16~ E_g$.
Note that we do not take into account any specific donor molecule, however, since $E_D > E_F$, i.e. there can be a movement of electrons from the donors to the semiconductor, and $E_F> E_{CB}$, i.e. the n-type semiconductor is heavily doped, our assumptions are sensible.
\begin{table}[h]
\caption{Parameters relative to the energy bands, used for the calculation of $n_0(\mathbf{r})$:}
\label{table:parameters}
\begin{ruledtabular}
\begin{tabular}{ccccc}
$T$~(K) &$E_{VB}$ & $E_{CB}$ & $E_D$& $N_D$~(cm$^{-3}$)\\
\hline    
$300$&  $0$ & $E_g$ & $1.40~ E_g$ & $10^{19}$\\
\end{tabular}
\end{ruledtabular}
\end{table}
\begin{table}[h]
\caption{InP properties}
\label{table:InP_parameters}
\begin{ruledtabular}
\begin{tabular}{ccccccc}
$m$ & $m_h$ & $\varepsilon$ & $\varepsilon_{\infty}$ & E$_\text{g}$~(eV)& $\gamma$~(ps$^{-1}$) & $\tilde{\lambda}_{\rm p}$~(\um)\\
\hline    
 $0.078~m_e$&  $0.089~m_e$ & $12.5$& $9.55$ & $1.344$ & $10$&$9.1$\\
\end{tabular}
\end{ruledtabular}
\end{table}
\newpage
\subsection{Nonlinear sources}
In the main text we have derived the following system to describe harmonic generation in heavily doped semiconductors in the case of a space variable equilibrium charge density $n_0 (\mathbf{r})$:

\begin{subequations}
\label{eqn:HM_set_SI}
    \begin{eqnarray}
        &&\nabla\times\nabla\times \mathbf{E}_j
        -\varepsilon\frac{\omega_j^2}{c^2}\mathbf{E}_j
        -\omega_1^2\mu_0 (\mathbf{P}_j+\mathbf{P}_{\omega_j}^{\rm NL})=0,    
        \label{eqn:maxwell}
\\
        &&\beta^2\nabla ({\nabla\cdot\mathbf{P}_j})
        -\frac{1}{3}\frac{\beta^2}{ n_0}(\nabla\cdot\mathbf{P}_j)\nabla n_0
        +(\omega^2+i\gamma\omega)\mathbf{P}_j
        = -\frac{n_0e^2}{m}\mathbf{E}_j +\mathbf{S}_{\omega_j}.
        \label{eqn:HM_harmonic_SI}
    \end{eqnarray}
\end{subequations}

Here, the nonlinear source terms due to free charges are: 

\begin{eqnarray}
        &&\mathbf{S}^{(2)}_{\omega_2}=-\frac{e}{m}(\mathbf{E}_1\nabla\cdot \mathbf{P}_1)
        -i\frac{e\mu_0}{m}\omega{\mathbf{P}_1}\times {\mathbf{H}_1}
        +\frac{\omega^2}{e n_0}({{\mathbf{P}_1}\nabla\cdot {\mathbf{P}_1} 
        +{\mathbf{P}_1}\cdot\nabla {\mathbf{P}_1}})
        -\frac{\omega^2}{e n_0^2}{\mathbf{P}_1}({\mathbf{P}_1}\cdot\nabla n_0) \nonumber
    \\
        &&\hspace{1.05 cm}
        +\frac{1}{9}\frac{\beta^2}{e n_0^2}(\nabla\cdot\mathbf{P}_1)^2\nabla n_0
        -\frac{1}{3}\frac{\beta^2}{e n_0} \nabla (\nabla\cdot\mathbf{P}_1)^2,
\end{eqnarray}

for the SHG and $\mathbf{S}_{\omega_3}=\mathbf{S}_{\omega_3}^{(2)}+\mathbf{S}_{\omega_3}^{(3)}$ for the THG, with :

\begin{subequations}
\label{eqn:NL_sources}
        \begin{eqnarray}
        &&\mathbf{S}^{(2)}_{\omega_3}=-\frac{e}{m}\left({\mathbf{E}_2}\nabla  \cdot {\mathbf{P}_1}+{\mathbf{E}_1}\nabla  \cdot {\mathbf{P}_2}\right)
        -i\frac{e\mu_0}{m}\left(\omega_2{\mathbf{P}_2} \times {\mathbf{H}_1}+\omega_1{\mathbf{P}_1} \times {\mathbf{H}_2}\right)
        \nonumber
    \\
        &&\hspace{1.05 cm}
        +\frac{\omega_1\omega_2}{e{n_0}}({{\mathbf{P}_2}\nabla \cdot {\mathbf{P}_1} + {\mathbf{P}_2}\cdot \nabla {\mathbf{P}_1} + {\mathbf{P}_1}\nabla \cdot {\mathbf{P}_2} + {\mathbf{P}_1}\cdot \nabla {\mathbf{P}_2}})
        -\frac{\omega_1\omega_2}{e n_0^2} \left[{\mathbf{P}_2}({\mathbf{P}_1}\cdot\nabla n_0)
        +{\mathbf{P}_1}({\mathbf{P}_2}\cdot\nabla n_0)\right]
        \nonumber
    \\
        &&\hspace{1.05 cm}
        -\frac{2}{3}\frac{\beta^2}{en_0}\left(\nabla \cdot {\mathbf{P}_2}\nabla \nabla  \cdot {\mathbf{P}_1}+\nabla  \cdot {\mathbf{P}_1}\nabla \nabla  \cdot {\mathbf{P}_2}\right)
        +\frac{2}{9}\frac{\beta^2}{e n_0^2}(\nabla\cdot\mathbf{P}_1)(\nabla\cdot\mathbf{P}_2)\nabla n_0, 
    \\
        &&\mathbf{S}_{\omega_3}^{(3)}=-\frac{\omega _1^2}{e^2n_0^2}
        \Big[
        \nabla \cdot \mathbf{P}_1 (\mathbf{P}_1\nabla  \cdot \mathbf{P}_1
        +\mathbf{P}_1 \cdot \nabla \mathbf{P}_1 )
        +\mathbf{P}_1\cdot \mathbf{P}_1\nabla\nabla\cdot\mathbf{P}_1
        \Big]\nonumber
    \\
        &&\hspace{1.05 cm}
        +\frac{2\omega^2}{e^2n_0^3}\Big[(\nabla\cdot \mathbf{P}_1){\mathbf{P}_1}( {\mathbf{P}_1}\cdot \nabla n_0)\Big]
        +\frac{1}{27}\frac{\beta^2}{e^2 n_0^2}\nabla (\nabla\cdot {\mathbf{P}_1})^3
        -\frac{4}{81}\frac{\beta^2}{e^2 n_0^3}(\nabla\cdot\mathbf{P}_1)^3\nabla n_0,
    \end{eqnarray}
\end{subequations}
describing cascaded and direct THG due to FE dynamics, respectively.
\end{widetext}

\end{document}